\documentclass[reprint,amsmath,amssymb,aps,prl,superscriptaddress]{revtex4-1}

\usepackage{graphicx}
\usepackage{dcolumn}
\usepackage{bm}
\usepackage[space]{grffile}
\usepackage[export]{adjustbox} 
\usepackage{booktabs}
\usepackage[caption=false]{subfig}
\usepackage{array}
\usepackage[flushleft]{threeparttable}
\usepackage{natbib}

\begin{document}
\title{Transforming mesoscale granular plasticity through particle shape}
\author{Kieran A. Murphy}
\affiliation{James Franck Institute, The University of Chicago, Chicago, IL 60637, USA}
\author{Karin A. Dahmen}
\affiliation{Department of Physics, University of Illinois at Urbana Champaign, Urbana, IL 61801, USA}
\author{Heinrich M. Jaeger}
\affiliation{James Franck Institute, The University of Chicago, Chicago, IL 60637, USA}

\date{\today}

\begin{abstract}

When an amorphous material is strained beyond the point of yielding it enters a state of continual reconfiguration via dissipative, avalanche-like slip events that relieve built-up local stress. 
However, how the statistics of such events depend on local interactions among the constituent units remains debated. 
To address this we perform experiments on granular material in which we use particle shape to vary the interactions systematically. 
Granular material, confined under constant pressure boundary conditions, is uniaxially compressed while stress is measured and internal rearrangements are imaged with x-rays. 
We introduce volatility, a quantity from economic theory, as a powerful new tool to quantify the magnitude of stress fluctuations, finding systematic, shape-dependent trends. 
In particular, packings of flatter, more oblate shapes exhibit more catastrophic plastic deformation events and thus higher volatility, while rounder and also prolate shapes produce lower volatility. 
For all 22 investigated shapes the magnitude $s$ of relaxation events is well-fit by a truncated power law distribution  $P(s)\sim {s}^{-\tau} exp(-s/s^*)$, as has been proposed within the context of plasticity models. 
The power law exponent $\tau$ for all shapes tested clusters around $\tau=$ 1.5, within experimental uncertainty covering the range 1.3 - 1.7. 
The shape independence of $\tau$ and its compatibility with mean field models indicate that the granularity of the system, but not particle shape, modifies the stress redistribution after a slip event away from that of continuum elasticity. 
Meanwhile, the characteristic maximum event size $s^*$ changes by two orders of magnitude and tracks the shape dependence of volatility. Particle shape in granular materials is therefore a powerful new factor influencing the distance at which an amorphous system operates from scale-free criticality. 
These experimental results are not captured by current models and suggest a need to reexamine the mechanisms driving mesoscale plastic deformation in amorphous systems. 

\end{abstract}

\maketitle

\section{Introduction}
Earthquakes, magnetic avalanches in Barkhausen noise, and sudden slip events during plastic deformation of a granular material all are examples of the complex dynamic response of a many-component system that is driven at fixed, slow rate. 
While individual events are unpredictable from one to the next, the statistics relating event magnitude and frequency exhibit remarkable similarity across a wide range of amorphous systems and size scales, including metallic glasses, emulsions, foams, granular materials, ice, as well as metals and alloys \cite{ArgonReview2013,DenisovDahmenSchall2016,BubbleRaftBMG,Ice,UniversalQuakeStats2015}. 
A common reference scenario for this dynamic response has been proximity to a non-equilibrium critical point, resulting in intermittent dynamics and power law distributions for the event sizes.

\begin{figure*}
	\centering
	\includegraphics[width=1.0\linewidth]{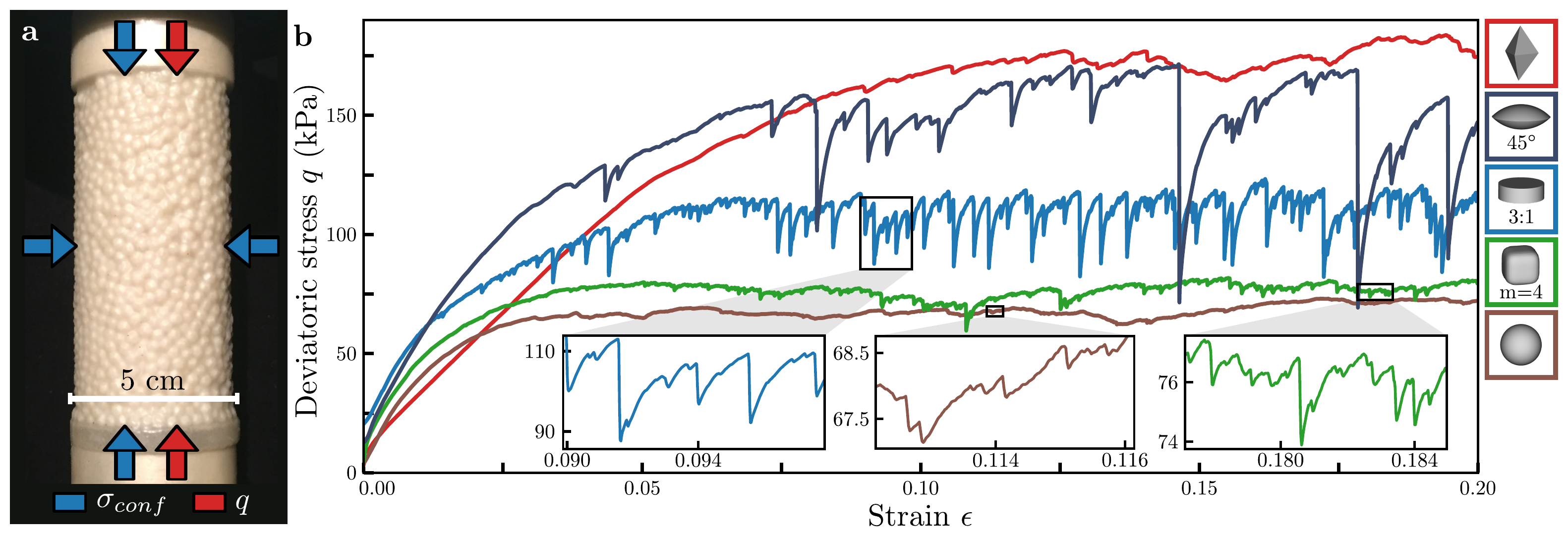}
	\caption{Triaxial compression into the plastic regime. (a) A 10cm tall, 5cm diameter packing of icosahedra in a rubber membrane and evacuated to $\sigma_{conf}=20$kPa before uniaxial stress $q$ is applied. (b) Raw stress-strain data for a single run each for the five particle shapes shown on the right (see Table 1 for shape definitions). All data in this paper are from the range $\epsilon=[0.1,0.2]$.}
\end{figure*}

The exponent $\tau$ of the power law describing the event magnitude distribution is important as it tells about the mechanism leading to scale-free dynamics. 
Mean field models predict $\tau$ = 3/2, independent of the details of the interactions among the system’s components \cite{FisherDepinning1998}. 
Simulations and experiments on plastic deformation of amorphous systems currently lack consensus or precision as to whether the exponent is indeed consistent with 3/2 or not \cite{DenisovDahmenSchall2016,UniversalQuakeStats2015,Salje2014,Antonaglia2014,RobbinsDamping2013,KDaniels2011}. 

On the modeling side, much of this can be traced back to differences in which the stress redistribution after a local plastic event is treated, i.e., to different treatments of the mesoscale dynamics \cite{Barrat2017arxiv}. 
When a relaxing volume element increases the stress on all neighboring elements, as happens in the presence of shear bands in, e.g., bulk metallic glasses and granular materials on larger scales, a mean field exponent $\tau$ = 3/2 is found \cite{DahmenSimpleAnalytic2011,Yip2018}. 
When, instead, stress redistribution takes the form of a quadrupolar kernel derived from Eshelby's work for localized plastic zones in elastic media \cite{Eshelby1957}, the exponent reduces to approximately $\tau$ = 1.3 in three-dimensional systems in the absence of slip localization \cite{Wyart2014, TensorModelZapperi2017}.

On the experimental side, discerning between $\tau=$ 1.3 and 1.5 requires several decades of events to be observed and careful analysis \cite{Antonaglia2014,ClausetPowerLaws2009,DahmenTimeResolution2016}. 
Most importantly, experimental investigation into the plasticity mechanisms operative at the mesoscale  has been scarce due to the difficulty of varying local interactions in systems whose constituent units are atoms, molecules, or even bubbles.  
Here we show how this can be achieved with a granular material by changing the particle shape. 

Granular materials present unique opportunities as model systems for the study of amorphous plasticity because their macroscopic scale allows for direct access to the parameters that govern local stress transmission and redistribution. 
The idea that particle shape is an important driver in a granular material's plastic behavior has been pursued in simulations \cite{AzemaPentagons2007,AzemaAngularity2012,AzemaPlaty2013}, but past experiments have either been limited to two-dimensional systems \cite{KDaniels2011, BaresPRE2017, Harrington2018} or confined to spheres \cite{Doanh2013, CuiTriax2016, OzbaySSTriax2016, DenisovDahmenSchall2016, GlassBeadsAngle2010} or naturally occurring soils and grains \cite{Santamarinasoilshape2004, AnthonyParticleCharacteristics2005, Marone2002, Flatparticleslayflat1974}. 
Our work takes advantage of 3D-printing in order to create particles whose shape-dependent interactions with contacting neighbors can be tailored with precision, while parameters such as the particles' material stiffness and their surface friction can be kept unchanged. 

Specifically, by varying particle shape we are able to change the manner with which particle surfaces meet to support stress, whether by edges, corners, or surfaces with different radii of curvature. Shape also drives where on each particle contacts are likely to occur, leading to different proportions of body forces versus torques. Finally, the resistance particles feel toward reconfiguring along rotational and translational degrees of freedom changes with particle shape (e.g., a flat disk would rather slide than rotate out of plane, and this preference strengthens as the aspect ratio of disks becomes larger). In these ways and others, how stresses are passed around at the microscale is modifiable through shape to a degree impossible in other plasticity experiments. Granular materials composed of different particle shapes are therefore a powerful new system with which to study 3D plasticity at the mesoscale.

To initiate plastic deformations, we perform uniaxial compression on columns comprised of several thousand copies of a chosen particle shape, randomly packed inside an elastic sleeve and subjected to a fixed confining pressure (Fig. 1a). 
As the applied strain is increased beyond an initial loading phase, the packing yields and eventually enters a regime referred to in soil mechanics as the critical state \cite{SchofieldCriticalState1968}. 
In this regime the stress has leveled off and fluctuates around a mean value as the packing restructures via nonaffine, dissipative particle rearrangements. 
After each of these reconfigurations, the material has locally self-healed and the column can load up again (Fig. 1b). 

We focus on mesoscale dynamics, where the length scale of the system is an order of magnitude larger than a characteristic rearrangement event (Fig. 2), but small enough to inhibit shear banding.  
When amorphous metal nanopillars are made at the mesoscale ($\sim$100nm in width), homogeneous deformation takes the place of shear localization and the result is a stronger material with desirable ductility \cite{VolkertBMGNano2008,Greer2010}. 
We are able to study the same physics at the centimeter scale, easily keeping the size of the granular column in the range of 10-100 particle diameters, but with the added capabilities of modifying the constituent particles of the system and directly imaging the individual rearrangement events with X-rays.  
Additionally, the mesoscale is precisely where amorphous plasticity theory struggles: in larger systems shear banding  is thought to make mean field approaches valid ($\tau=1.5$), while in smaller systems without shear banding  correlations can arise from Eshelby stress redistribution and lead to $\tau\sim1.3$. As a result, mesoscale granular systems without shear bands offer an excellent testbed for studying the intermittent dynamics of plastic deformation in amorphous materials.

Figure 1b shows representative data from 5 of the 22 different particle types investigated (see Table 1). 
As the applied strain is ramped up, both gradual stress variations and sudden stress drops due to near instantaneous relaxation events are found. 
Note the widely differing magnitude and character of the fluctuations around the average stress level, despite the fact that all samples were prepared and measured under identical conditions. 

In order to quantify the magnitude of sudden stress fluctuations, we introduce a measure borrowed from financial mathematics which quantifies the spread of fractional changes that occur in a time series. 
This measure, volatility, is model-independent and is particularly useful in comparing broadly distributed fluctuations in data with different or changing baselines \cite{InvestmentBook}.
Importantly, it sidesteps the issue faced in many plasticity experiments, of accurately locating and measuring the magnitude of stress drops\cite{DahmenTimeResolution2016,DahmenLowRes2016}, giving a robust method to quantify the jerkiness of plasticity data across experiments and even across systems.
Plotting volatility versus the angle of internal friction, a measure of a granular material's shear strength \cite{SchofieldCriticalState1968}, then enables us to extract trends in the way particle geometry correlates with strength and fluctuations in the plastic regime. 

\begin{table}[ht]
 \centering 
 \begin{tabular}{c c c c} 
 \hline\hline\\ 
 Shape & Symbol & Shape & Symbol\\ [0.5ex] 
 \hline\\

	Sphere & \parbox[c]{0.2in}{
  \includegraphics[width=0.2in]{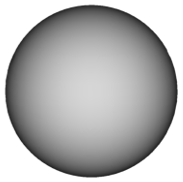}} & Lens, $\gamma=75^{\circ}$ & \parbox[c]{0.25in}{
  \includegraphics[width=0.25in]{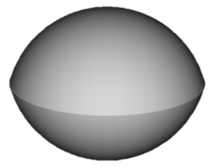}} \\ 
  Tetrahedron & \parbox[c]{0.2in}{
  \includegraphics[width=0.2in]{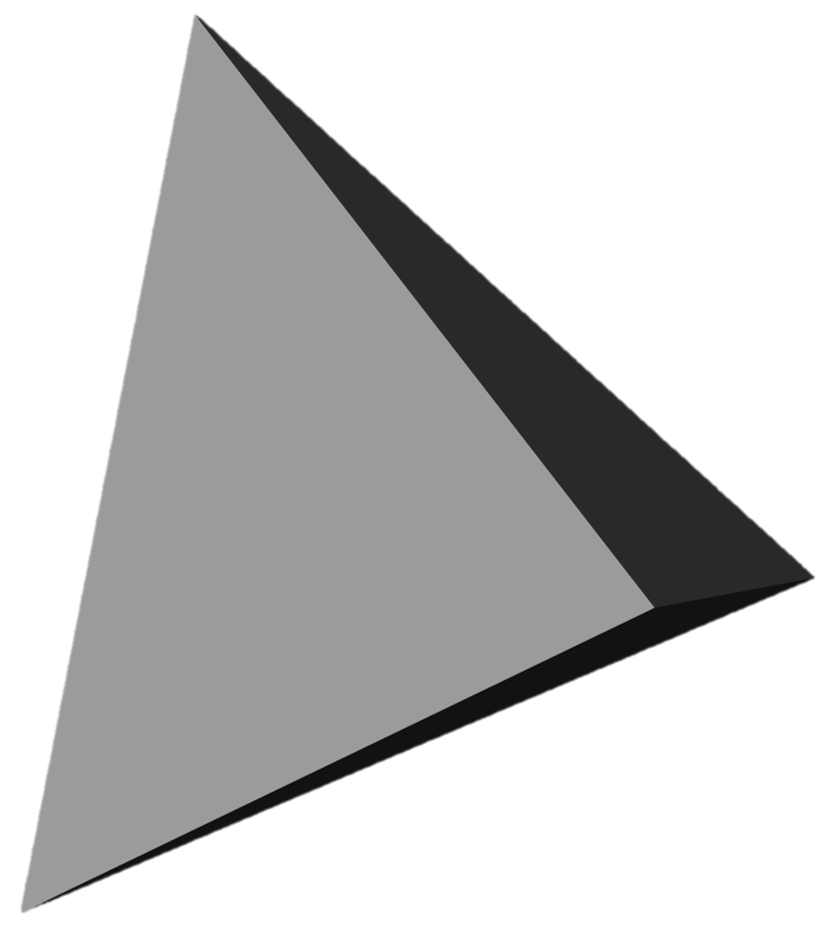}} & Lens, $\gamma=60^{\circ}$ & \parbox[c]{0.25in}{
  \includegraphics[width=0.25in]{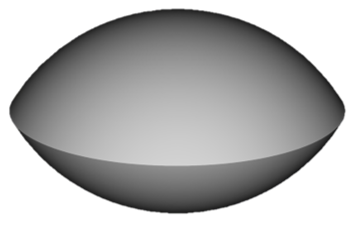}} \\
  Triangular bipyramid & \parbox[c]{0.2in}{
  \includegraphics[width=0.2in]{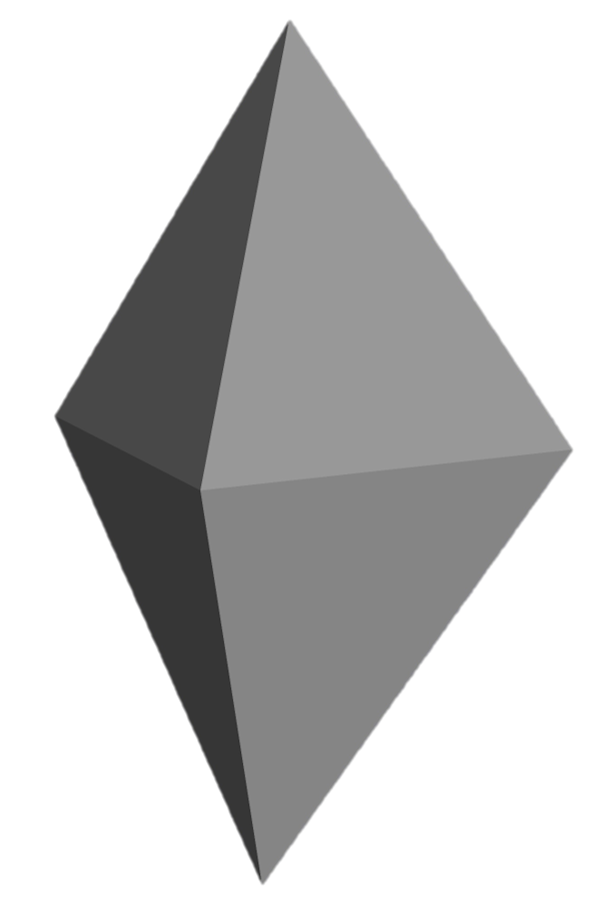}} & Lens, $\gamma=52.5^{\circ}$ & \parbox[c]{0.29in}{
  \includegraphics[width=0.29in]{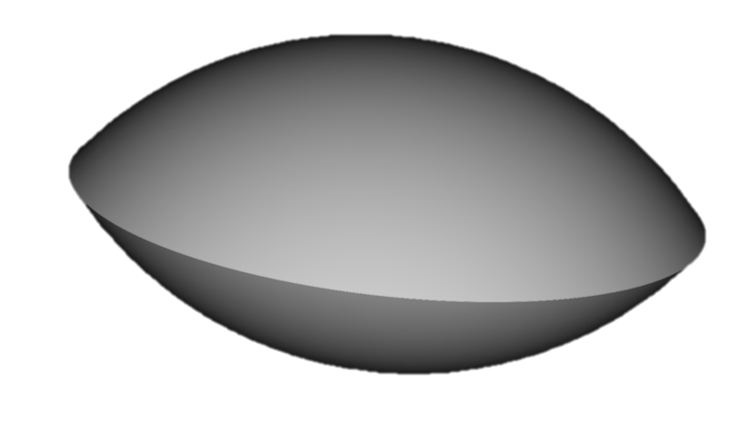}} \\
  Cube & \parbox[c]{0.22in}{
  \includegraphics[width=0.22in]{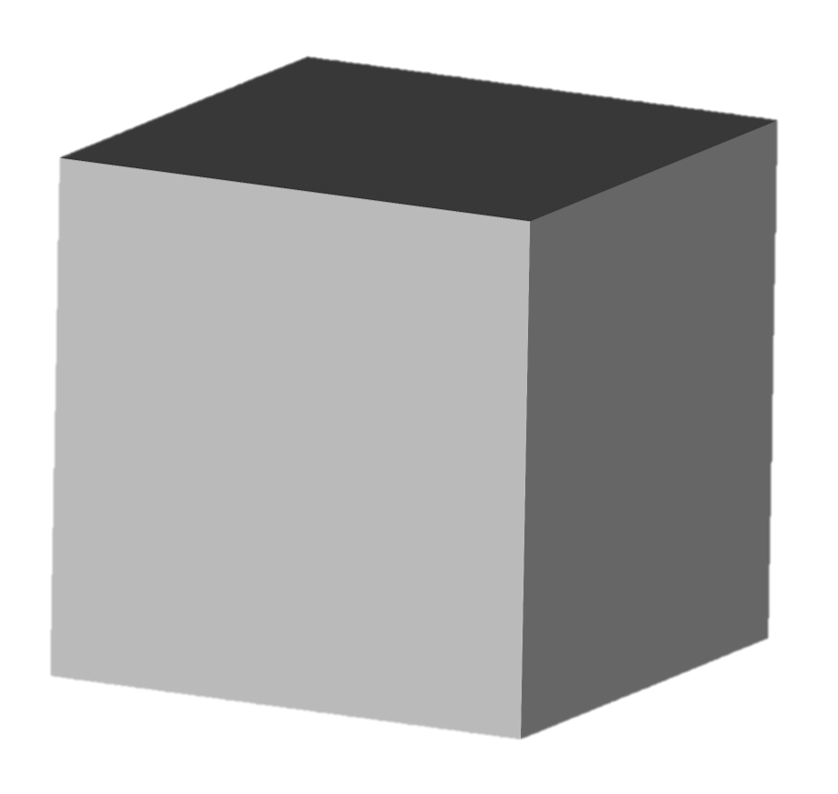}} & Lens, $\gamma=45^{\circ}$ & \parbox[c]{0.28in}{
  \includegraphics[width=0.28in]{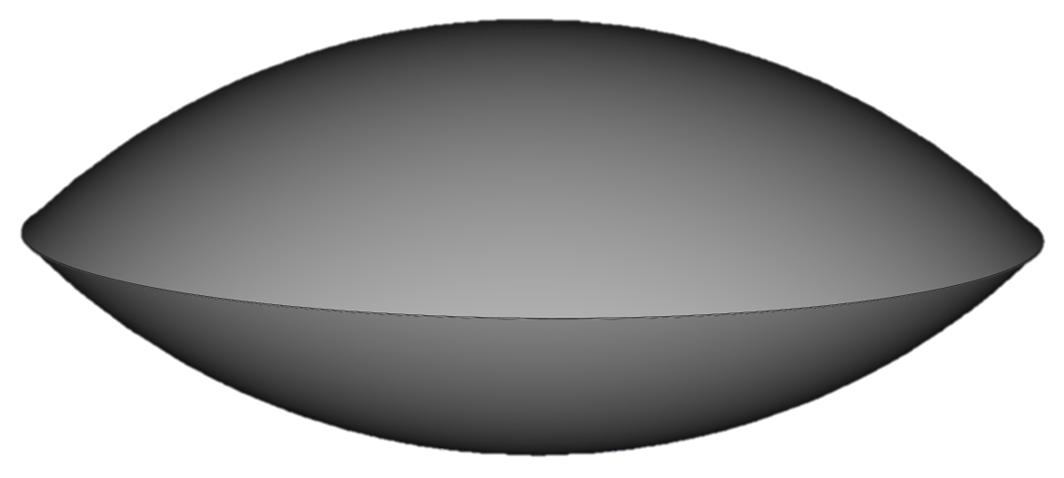}} \\
  Octahedron & \parbox[c]{0.24in}{
  \includegraphics[width=0.24in]{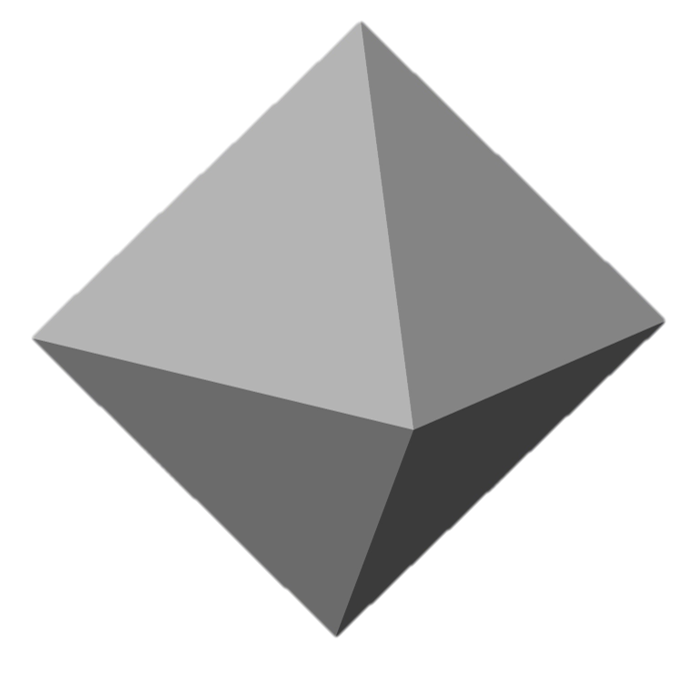}} & Cone, $\beta=45^{\circ}$ & \parbox[c]{0.23in}{
  \includegraphics[width=0.23in]{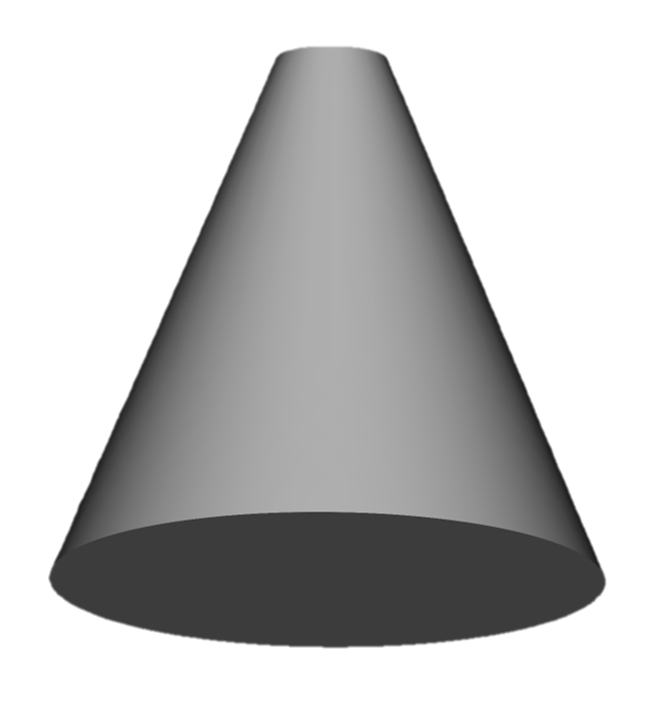}}\\
  Dodecahedron & \parbox[c]{0.29in}{
  \includegraphics[width=0.24in]{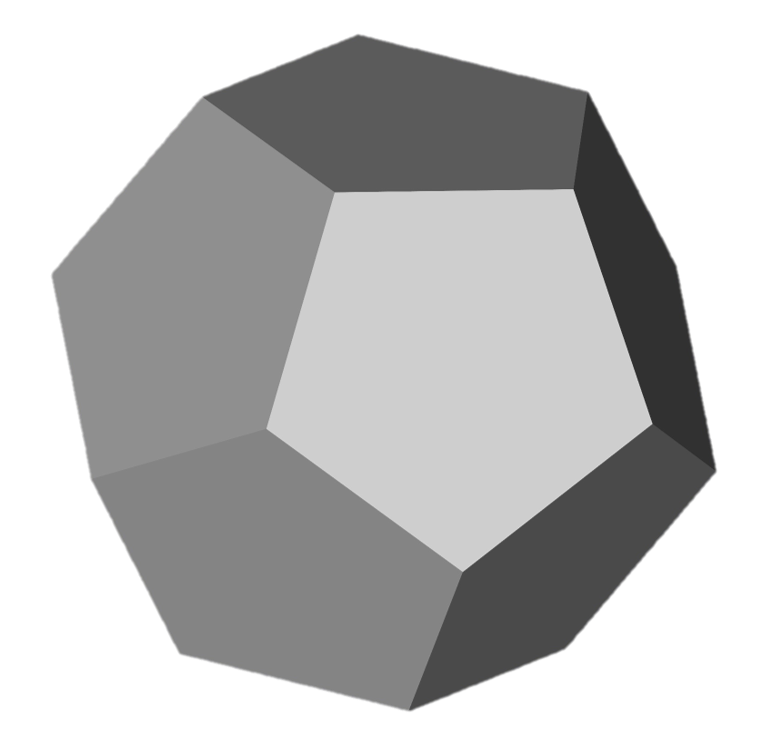}} & Disk, $D/t=1.5$ & \parbox[c]{0.2in}{
  \includegraphics[width=0.2in]{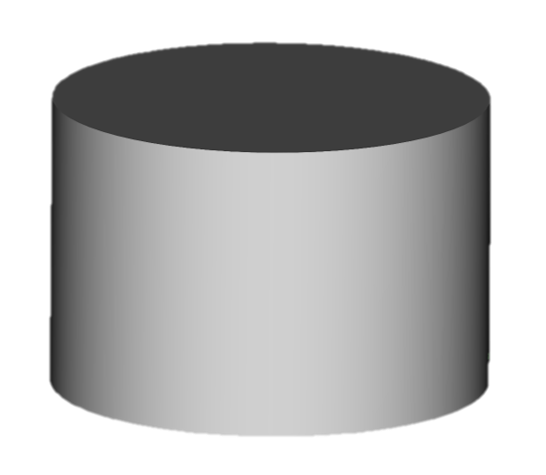}} \\ 
  Icosahedron & \parbox[c]{0.2in}{
  \includegraphics[width=0.2in]{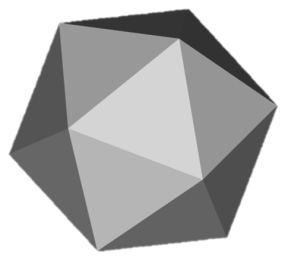}} & Disk, $D/t=3$ & \parbox[c]{0.2in}{
  \includegraphics[width=0.2in]{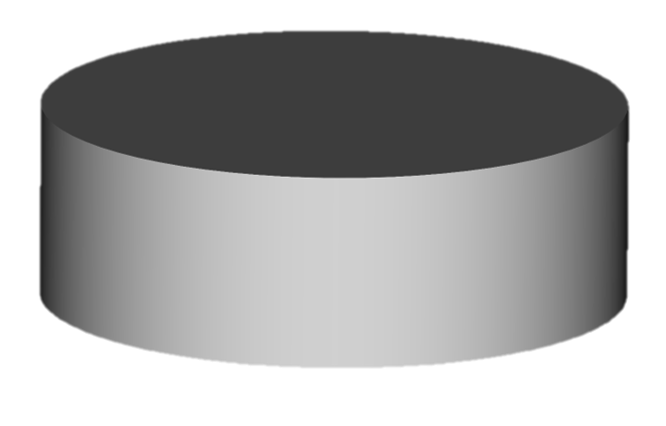}}\\
  Supercube, $m=3$ & \parbox[c]{0.22in}{
  \includegraphics[width=0.22in]{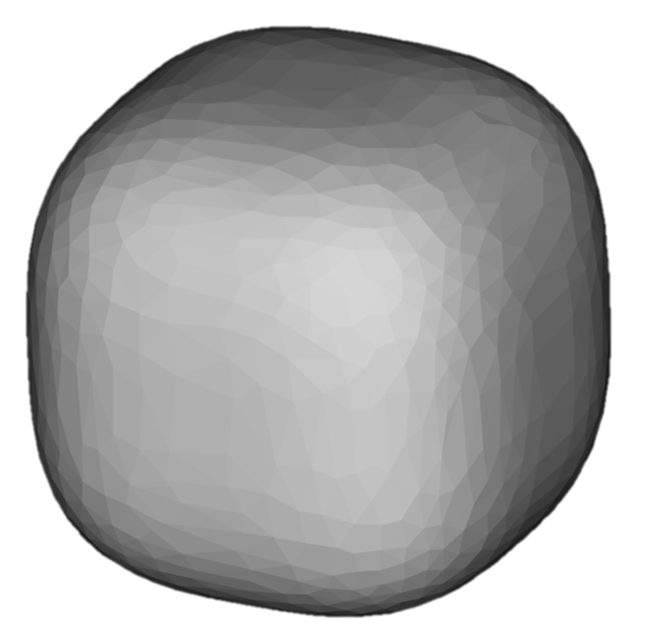}} & Disk, $D/t=4$ & \parbox[c]{0.2in}{
  \includegraphics[width=0.2in]{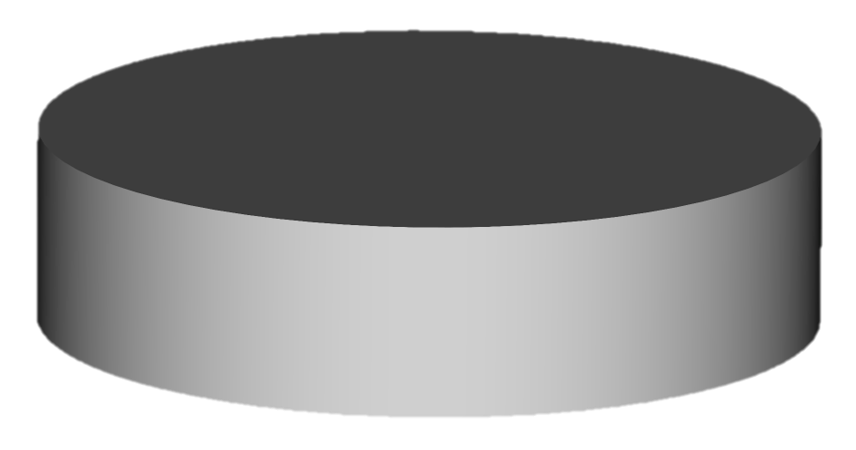}}\\
  Supercube, $m=4$ & \parbox[c]{0.22in}{
  \includegraphics[width=0.22in]{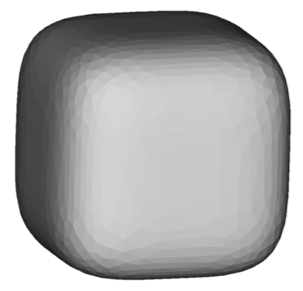}} & Disk, $D/t=4.5$ & \parbox[c]{0.2in}{
  \includegraphics[width=0.2in]{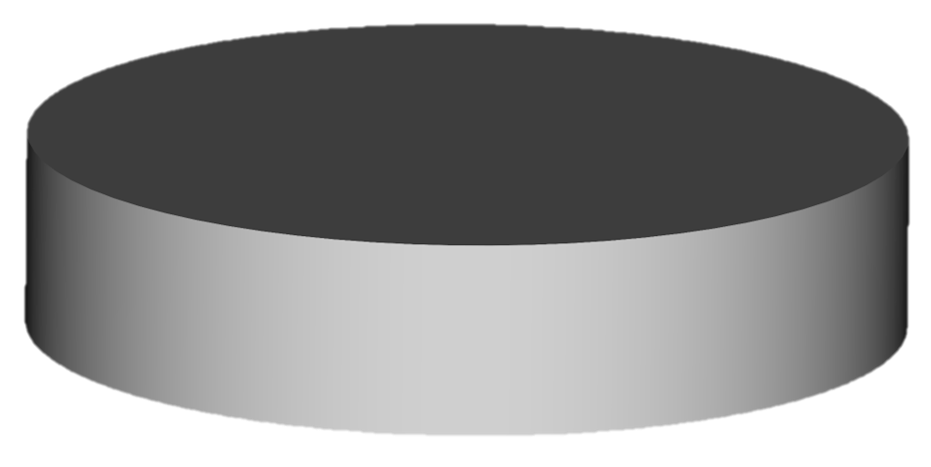}} \\
  Divot cube & \parbox[c]{0.25in}{
  \includegraphics[width=0.25in]{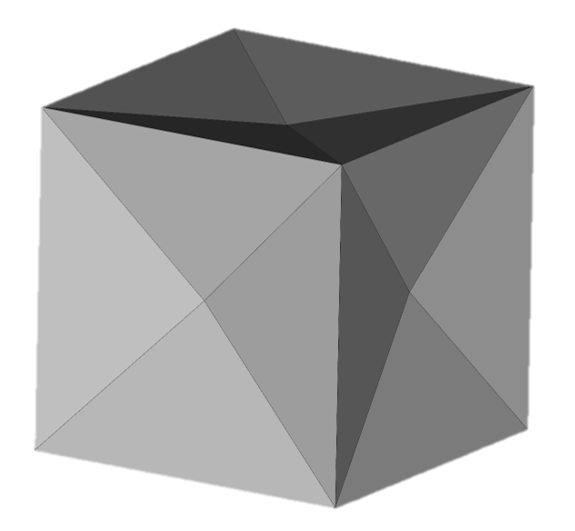}} & Corner & \parbox[c]{0.21in}{
  \includegraphics[width=0.21in]{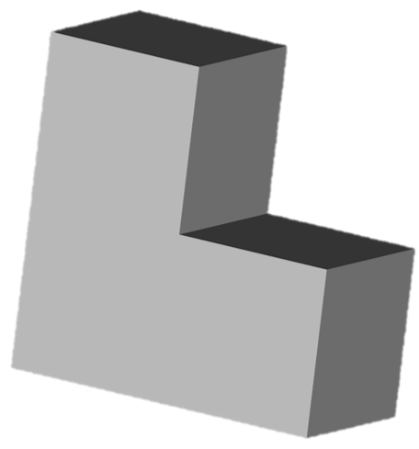}} \\
  Hemisphere & \parbox[c]{0.25in}{
  \includegraphics[width=0.25in]{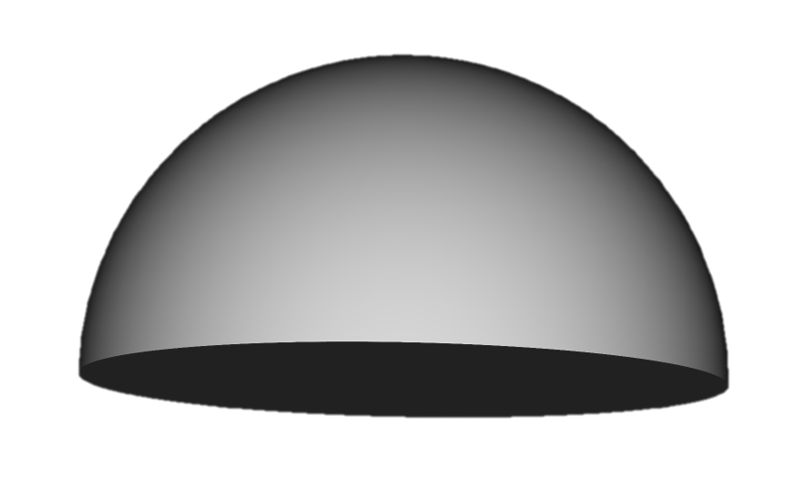}} & Prolate ellipsoid, 2.5:1:1 & \parbox[c]{0.25in}{
  \includegraphics[width=0.12in]{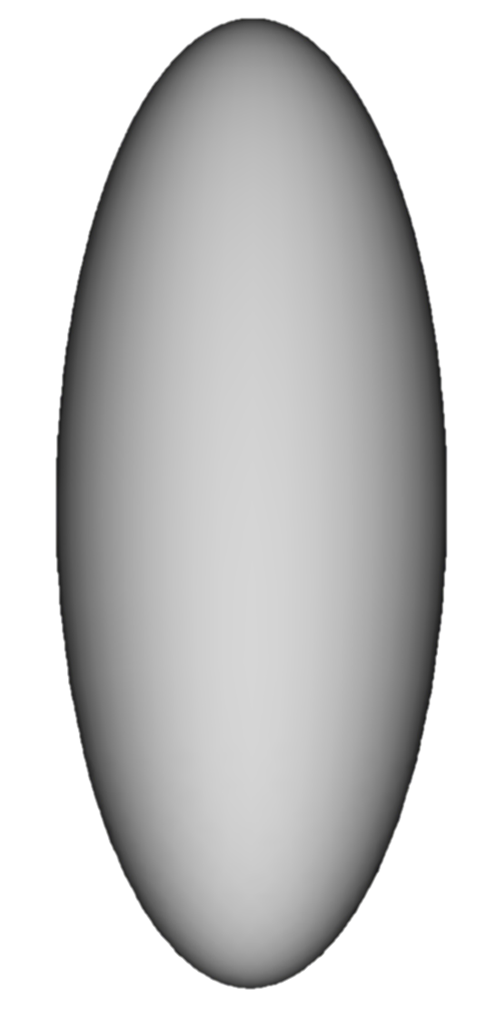}} \\
  [1ex] 
  \hline 
  \end{tabular}
  \caption{Nonstandard shapes: A triangular bipyramid is two tetrahedra joined face to face. 
  A supercube surface is defined by $|x|^m+|y|^m+|z|^m=R^m$, with $m$=2 corresponding to a sphere and $m$=$\infty$ a cube.
  A divot cube has a right pyramid removed from each face such that edges sharpen from $90^\circ$ to $71^\circ$. A lens particle is the union of two spherical caps with polar angle $\gamma$; a lens particle with $\gamma=90^\circ$ is therefore a sphere. 
  The angle $\beta$ is the aperture of a cone particle; the tip of the cone is removed to prevent puncture of the latex membrane so the more precise shape is a frustum with $r_1/r_2=5$. A corner is three adjoined cube particles.}
	\label{table:shapes12} 
\end{table}

Similar to what has been reported for other plastically deforming systems \cite{UniversalQuakeStats2015,DenisovDahmenSchall2016}, we find that the shape of the distribution $D(s)$ of normalized drop magnitudes $s$ is well-fit, for all 22 particle shapes tested, by a truncated power law, $D(s)\sim s^{-\tau} exp(-s/s^*)$. 
While the power law exponent $\tau$ is found to always lie in a small neighborhood around 1.5, the cutoff stress scale $s^*$ is highly shape dependent. We discuss the general implications of this behavior in light of recent simulations and mesoscale plasticity models \cite{Barrat2017arxiv} and also the opportunities this opens up for designing specific stress responses with granular materials in the regime beyond yielding.

\begin{figure*}
	\centering
	\includegraphics[width=1.0\linewidth]{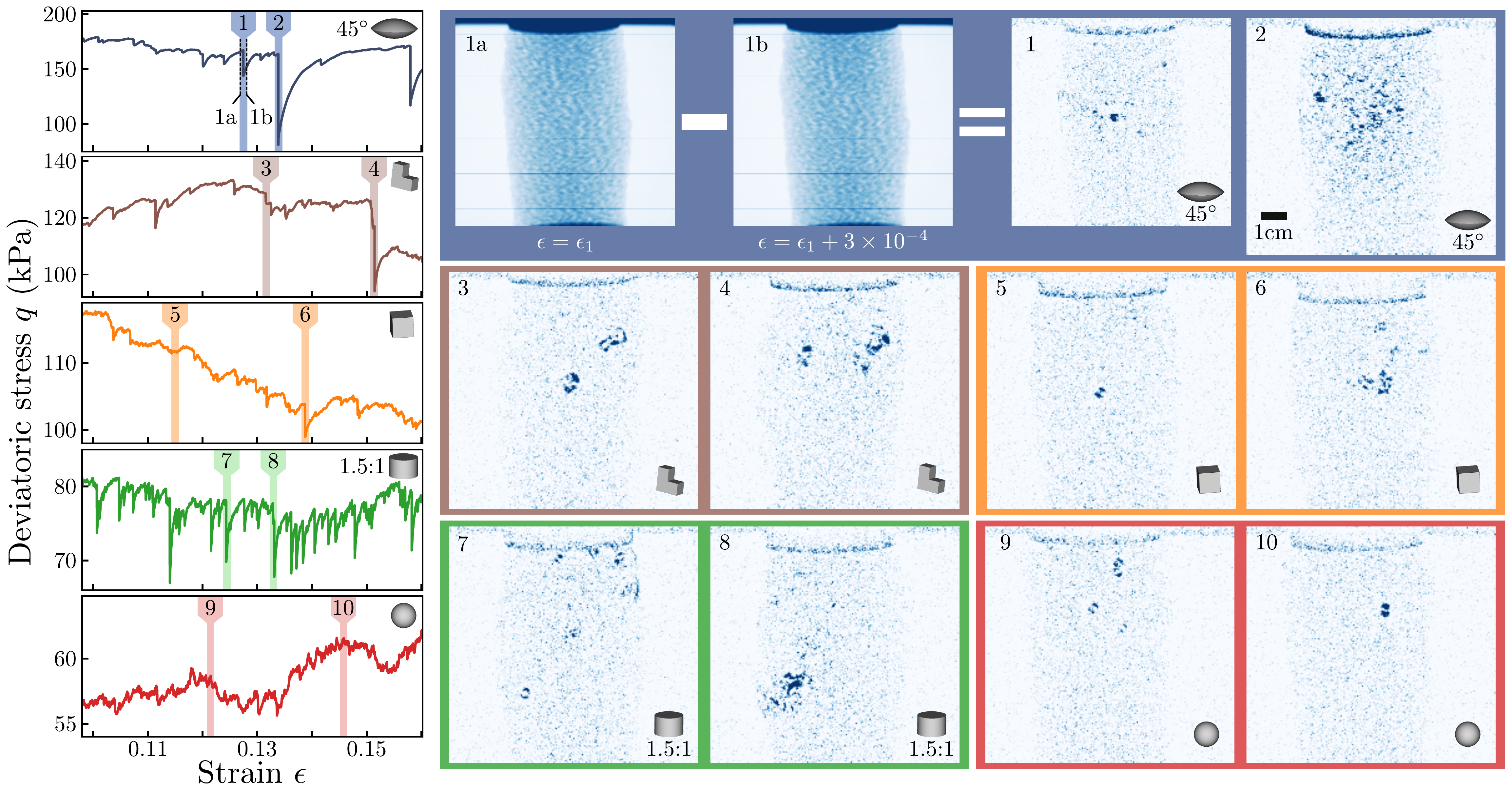}
	\caption{Localized slips revealed by X-ray videography. 
	Stress-strain data (left) for representative particle shapes, with numbered vertical bars that indicate the strains at which the X-ray data shown on the right were taken. Subtracting successive X-ray frames reveals the particle rearrangements that cause sudden stress drops in $q(\epsilon)$. We show this explicitly for the first labeled stress drop event. The dark horizontal band at the top of each X-ray difference image reflects the top plate's motion during compression. For a sense of scale, the dark blue regions in the difference images for events 5 and 10 each correspond to a single particle shifting position by roughly a particle diameter. For clarity the shaded bars in the stress-strain curves are plotted five times wider than the strain window captured by a single X-ray frame. The X-ray images for events 2-10 are all shown at the same scale, enabling direct comparison.}
\end{figure*}

\section{Experimental Details}
Table 1 lists the particle shapes used in our experiments. 
All particles were 3D-printed from UV-cured hard plastic (Young's modulus $E_{mat} \sim 1$GPa) using an Objet Connex350 printer with resolution $30\mu$m. 
Each particle's volume was 22.5mm$^3$, except for the corner particles, a shape comprised of three adjoined cubes making the volume 67.5mm$^3$. 
Except for the strain rate dependence data below, we controlled for surface properties of the particles by only using particle sets fresh out of the 3D-printer.

For each experiment about 5,000 particles (a third of that for the corner particles) were poured randomly into a 0.6mm thick, 5.0cm diameter latex membrane to form a column with aspect ratio 2:1 (height to diameter; Fig. 1a). 
A pump evacuated the interior of the column to apply an isotropic confining pressure $\sigma_{conf} = 20$kPa across the membrane. 
At fixed $\sigma_{conf}$ uniaxial compression tests were performed on an Instron 5800 series materials tester under strain-controlled loading conditions. 
The majority of the data presented here were taken with $\dot{\epsilon}=8\times10^{-4}$ s$^{-1}$ (0.05 min$^{-1}$), where the strain $\epsilon$ is the fractional axial displacement relative to the uncompressed column height. 

  The loading mode in compression experiments depends on $\alpha$, the ratio of the stiffness of the measurement apparatus to that of the sample \cite{Ghoniem2016}.
  The loading mode dictates how the apparatus responds immediately after a relaxation event. 
  In order to fully resolve individual stress drops of all sizes, large apparatus stiffness ($\alpha >>1$, strain control) and sufficiently slow compression ($\dot{\epsilon}\rightarrow0$, quasistatic limit) are required \cite{RouxStiffness2011}. In our experiments $\alpha \approx 3\times10^6 $Nm$^{-1}/3\times10^5$Nm$^{-1} =10$. A feature of this large $\alpha$ limit is clearly seen in Fig. 1b as the nonlinear recharge of stress after events, which shows how the granular packing is reloading. 
  This differs from typical stick-slip experiments, where $\alpha <$ 1 and the reloading is linear. The distinction is important: in the stick-slip case, the apparatus compresses during stick and surges forward into the sample during slip, using its own stored energy to fuel plastic events. 
  In the large $\alpha$ limit, plastic events run on the energy stored in the sample and thus the measured stress drops are more transparent indicators of the internal rearrangement process.

Data were taken at a rate of 40/s with 0.01N precision, corresponding to 5Pa in stress. 
At least five independent compressions were performed per particle shape in Figures 3-6, i.e., between each run the particles were poured out before starting the process again.
Some of the slow compressions in Fig. 7 were run only three times. 
All data discussed in this paper are from strains in the critical state past yielding, operationally associated here with $0.1\leq\epsilon\leq0.2$. 

Without additional processing beyond a simple threshold on the first derivative of the stress data, identification of small stress drops that occur during a recharge event will be missed or skewed to artificially smaller magnitudes.  
To avoid this, we pass the first derivative of the stress data $\Delta q/\Delta\epsilon$ through a high pass filter to remove baseline drift, similar to the method used in Ref. \cite{DahmenTimeResolution2016} to correct for low time resolution data. 
Specifically, we subtract from $\Delta q/\Delta\epsilon$ a Gaussian-smoothed version of itself, where the strain scale of the smoothing is the approximate duration of an event. 
An artifact of the filtering is that it lessens the magnitude of all events by an amount equal to the Gaussian smoothed version of the event; however, this is reversible since all drops in these experiments last the same amount of time -- the 50ms the apparatus takes to respond.  
The entire method was checked against and found to accurately recover drop distributions in synthetic time series data mimicking those in Fig. 1b.

To obtain information on the spatial extent of structural rearrangements, X-ray radiographs were taken with an Orthoscan C-arm fluoroscope at fixed intervals (2s on, 1.5s off) during several of the compression tests. 
These were run at a lower strain rate ($\dot{\epsilon}=8\times 10^{-5}$ s$^{-1}$) to accommodate the 2s image acquisition time. 
Successive X-ray frames were differenced to subtract off slow global deformations of the packing and highlight sudden localized particle rearrangements. 
The slow compression rate nicely separated these timescales, enhancing the difference images. 

\section{Results \& Discussion}
\subsection{Shape Dependent Features of Plastic Deformation}

The wide range of stress fluctuation behavior that emerges from varying particle shape is immediately apparent in the raw stress-strain data (Fig. 1b). 
Under compression all packings plastically ``flow'', but with very, and often surprisingly, different character. 
For example, with disk-shaped particles the behavior is strikingly intermittent, while triangular bipyramids with their sharp corners exhibit much smaller stress variations. 
The smaller fluctuations (e.g., for the spheres) appear comparatively smooth and continuous when plotted on the same scale as those of the disks (main panel of Fig. 1b), but on closer inspection can be as abrupt (see insets). 
X-ray imaging links the stress drops to particle movements, as shown in Fig. 2 for a selection of typical behaviors: small events are tied to the sudden movement of a single particle, clearly visible in events 5 and 10, while slightly larger drops may involve the rearrangement of a handful of particles in a local neighborhood, as seen in events 3 and 8. 
The most significant stress drops accompany restructuring that spans the column, as in events 2, 4 and 7. 
Interestingly, the rearrangements appear to be less localized than what would be expected in a shear transformation picture of plastic deformation \cite{ArgonShearTrans1979}. 
For instance, events 3 and 9 seem to involve 2-3 particles each, separated by several particle lengths where nothing moved noticeably. 
Such distributed local failure events are more in line with mesoscale models \cite{Wyart2014,TensorModelZapperi2017} and mean field treatments of granular plasticity\cite{DahmenSimpleAnalytic2011} where elastic interactions are taken to be long ranged.

To quantify and compare the degree by which stress-strain data as in Fig. 1b exhibit sudden jumps, we require a measure that is independent of the often large differences in mean stress, that can encapsulate the extremely wide range of observed fluctuation magnitudes, and that can ignore slowly drifting stress base lines. 
To this end we use the standard deviation of the logarithm of fractional changes, termed the volatility in the Black-Scholes model for price evolution \cite{InvestmentBook}. 
It provides a dimensionless, baseline-independent measure of the rate of change of a discretized time series. 
Treating the stress data as a series $q_i$, where each ``time'' step $i$ corresponds to an applied strain value, we define the instantaneous ``return'' 
\begin{equation}
R_i = ln(q_i/q_{i-1}).
\end{equation}
The volatility of the series is the sample standard deviation of $R$, given by
\begin{equation}
V = A\sqrt{\frac{1}{N-1} \sum_i^{N} (R_i-\bar{R})^2},
\end{equation}
with $N$ the number of strain intervals in the dataset and $A$ a constant to correct for the effect of data resolution. Finer resolution yields smaller fractional changes in $q$ and therefore a smaller value for $V$ if not corrected. We ``annualize'', as done in finance, with the constant $A$ equal to $\sqrt{\epsilon_{win}/\Delta\epsilon}$, the square root of the number of data points taken in a strain window $\epsilon_{win}$ relevant for analysis. For our data, the strain interval $\Delta\epsilon$ between measurements $q_{i-1}$ and $q_i$ is set by the sampling rate of the apparatus and strain rate of compression. To capture the mix of fluctuation behaviors in data as in Fig. 1b, we chose $\epsilon_{win}$=0.01.

Whereas poor time resolution can interfere with the measurement of avalanche distributions, the assumptions underlying the calculation of volatility are actually more applicable when the data acquisition timescale is slower than the timescale of a stress drop (but faster than the timescale separating drops). 
Specifically, the annualization rescaling relies on jumps in the time series being uncorrelated, which is invalid if the sampling rate is high enough to capture several data points during a stress drop. 
The model-independent nature of volatility and its applicability to low resolution data make it a broadly applicable tool for quantifying the jerkiness of time series data. 
Additionally, to a first order approximation, the return $R_i\sim \Delta q_i/q_i$ is the fractional change between timesteps, so the volatility can be interpreted as a spread in the fractional changes over the annualization time interval. 
In other words, a volatility value of $10^{-2}$ means stress fluctuations are on the scale of 1$\%$ of the average stress in each $\epsilon_{win}=0.01$ strain interval. 
Parenthetically, the volatility values for many common stock market indices fall in the neighborhood of 10\%, a comparable value to that of many of the more volatile shapes shown in Fig. 3.

In Fig. 3 we correlate the volatility $V$ with the average shear strength of the granular packing, as measured by the angle of internal friction $\psi$. 
This angle arises from the Mohr-Coulomb failure criterion in soil mechanics, and in \cite{SchofieldCriticalState1968} is defined for cohesionless grains through 
\begin{equation}
sin\psi = \frac{\sigma_3 - \sigma_1}{\sigma_3 + \sigma_1} = \frac{\bar{q}}{\bar{q}+2\sigma_{conf}}
\end{equation}
Here $\sigma_3$ and $\sigma_1$ are the largest and smallest principal stress components, equal to $\bar{q}+\sigma_{conf}$ and $\sigma_{conf}$ respectively, and $\bar{q}$ is the average deviatoric stress in the critical state. 
Extracting $\bar{q}$ from measurements of $q(\epsilon)$ involves subtlety if stress drops become so large that they correspond to a sizable fraction of the stress plateau reached in the critical state: in these cases a simple average of $q(\epsilon)$ tends to lie below this plateau stress and underestimates the stress needed for plastic deformation. 
To account for this, we instead define the plastic stress as the average of all of the stress values immediately preceding a stress drop, similar to what was done in \cite{CuiTriax2016}.

\begin{figure}
	\centering
	\includegraphics[width=1.0\linewidth]{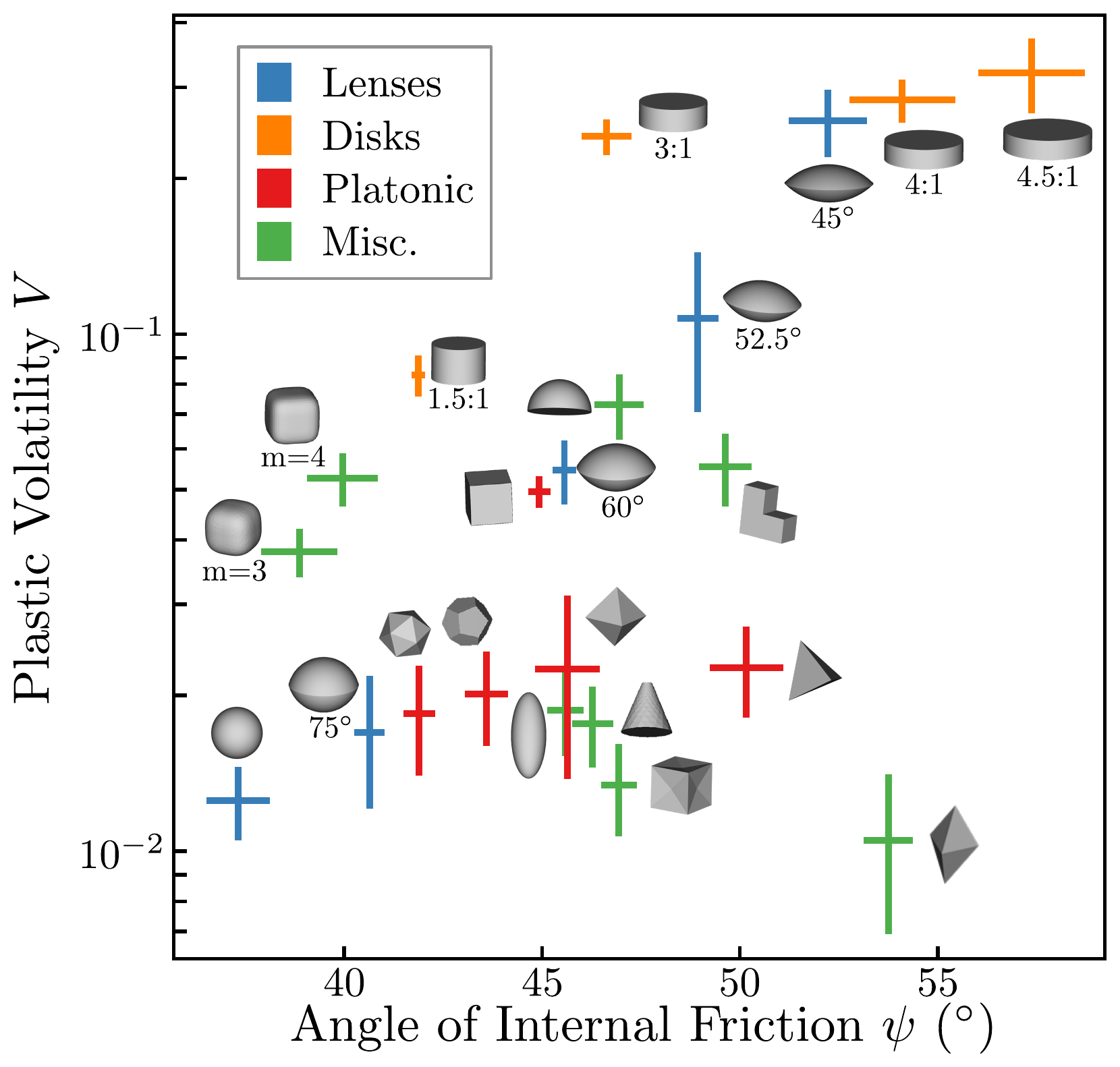}
	\caption{Plastic deformation phase space. 
    The jerkiness of the plastic regime and the shear strength are mapped out by plotting the annualized plastic volatility $V$ versus the angle of internal friction $\psi$. 
    Cross bars for each particle shape are the standard deviation of $V$ and $\psi$ across runs.}
\end{figure} 

\begin{figure}
	\centering
	\includegraphics[width=1.0\linewidth]{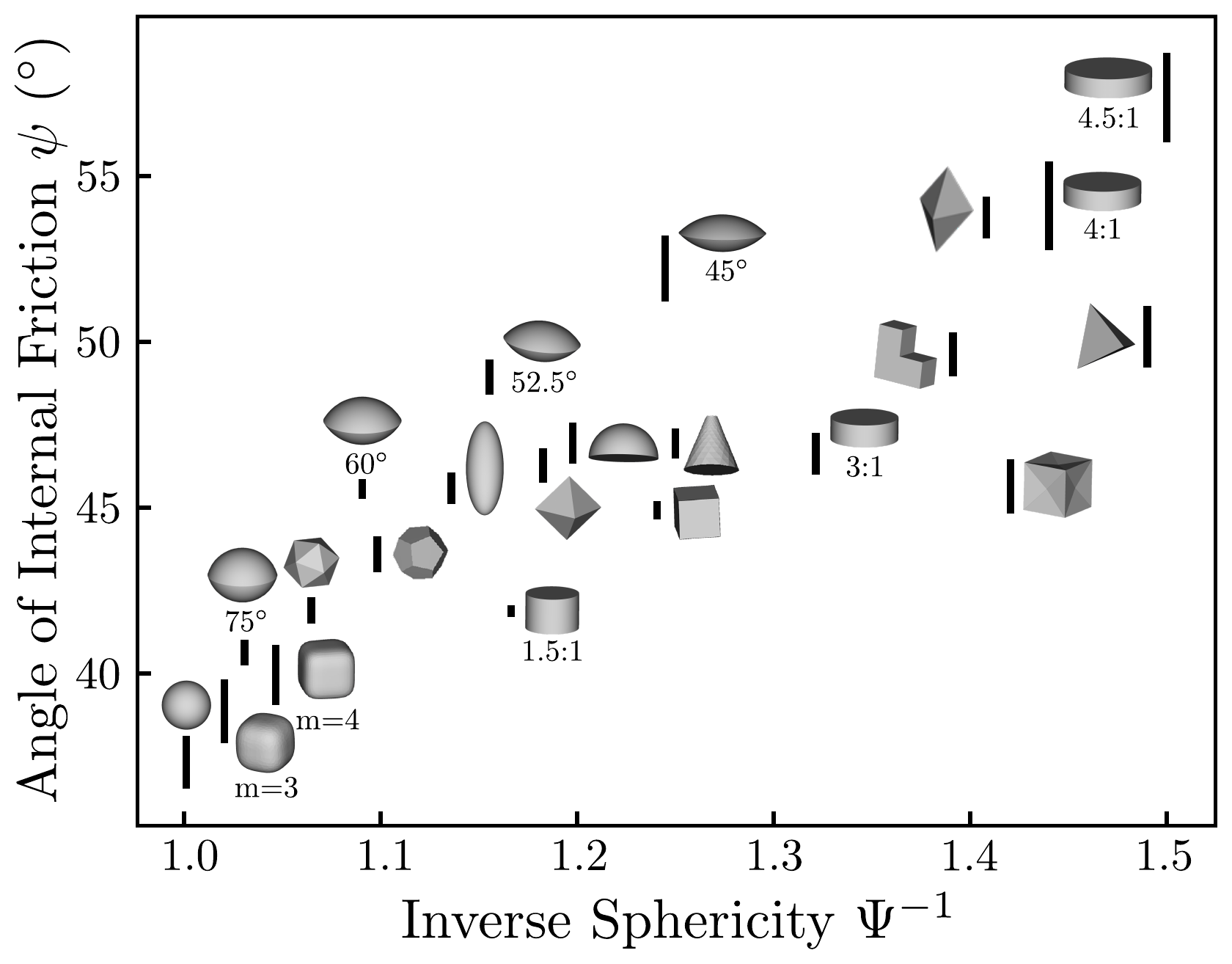}
	\caption{Shear strength versus deviations from a sphere. 
    The angle of internal friction data from Fig. 3 is plotted against the inverse of sphericity. 
    As a measure of the compactness, sphericity is calculated as the ratio of a particle's surface area to that of a sphere with the same volume.}
\end{figure} 

Figure 3 reveals several trends. Packings composed of spheres offer the least resistance to shear and thus exhibit low $\psi$ as well as low volatility. 
As $\psi$ increases, the region of nearly spherical particles splits into two branches, one containing more platy and oblate shapes and one more angular and prolate ones. 
These branches differ by more than an order of magnitude in $V$ at large $\psi$. 
As the set of lens-shaped particles demonstrates, both volatility and angle of internal friction are increased quickly when the lenses become more oblate. 
Disks follow a similar trend with increasing $\psi$ and $V$ as their aspect ratio (diameter to height) increases. 

Packings of platonic solids get stronger as the number of particle faces decreases from icosahedra to tetrahedra and as particle corners and edges become more pronounced. However, they generally exhibit quiescent deformation with $V$ values not much larger than those of spheres. 
Cubes are the exception with a roughly fourfold enhancement in $V$, providing a first hint about the relative importance of edges and faces. 
We explored this by altering the cubes: while rounding the edges into supercubes drops $\psi$ and fusing three cubes into a corner-shaped particle with larger faces and longer edges increases $\psi$, neither has a significant effect on $V$. 
On the other hand, while indenting flat cube faces by creating divots enhances $\psi$ slightly (presumably due to the sharper edges), it significantly reduces $V$, all the way down to the level of spheres.

These findings suggest the presence of competing factors: while shape anisotropy (oblate or prolate) as well as sharp edges or corners are all seen to enhance $\psi$, they do not predict $V$. 
Instead, $V$ appears to be more dependent on the degree to which particle contacts involve surfaces with large radius of curvature, such as faces rather than edges. 
This depends not only the existence of faces but also on the frequency of face-face contacts.
With the divot cubes we eliminated flat cube faces, while tetrahedra, a shape with large flat faces, is an example of a case where the packing structure does not favor face-face contacts \cite{SchroterTets2013}. Interestingly, the contacting areas do not have to be flat: increasing the radius of curvature at the contact enhances $V$. 
This is demonstrated by the set of lenses and also by comparing hemispheres with cones. The latter two shapes have very similar $\psi$ yet differ in the radius of curvature at contact, which mimics the trend in $V$.

In Figure 3 the angle of internal friction $\psi$ is a property of the packing as a whole. 
We can relate $\psi$ more directly to the geometry of individual particles as shown in Fig. 4, where we plot $\psi$ as a function of the inverse particle sphericity $\Psi^{-1}$. 
Sphericity is a measure of the compactness of a shape relative to a sphere (whose $\Psi$=1). 
As the data indicate, $\Psi^{-1}$ correlates reasonably well with $\psi$, as would other single-parameter shape descriptors, such as particle roundness \cite{Santamarinasoilshape2004} or isoperimetric quotient \cite{Glotzer2012}, highlighting the particles' resistance to rotation as the dominant driver in a granular packing's shear strength\cite{Marone2002}.

\begin{figure*}
	\centering
	\includegraphics[width=1.0\linewidth]{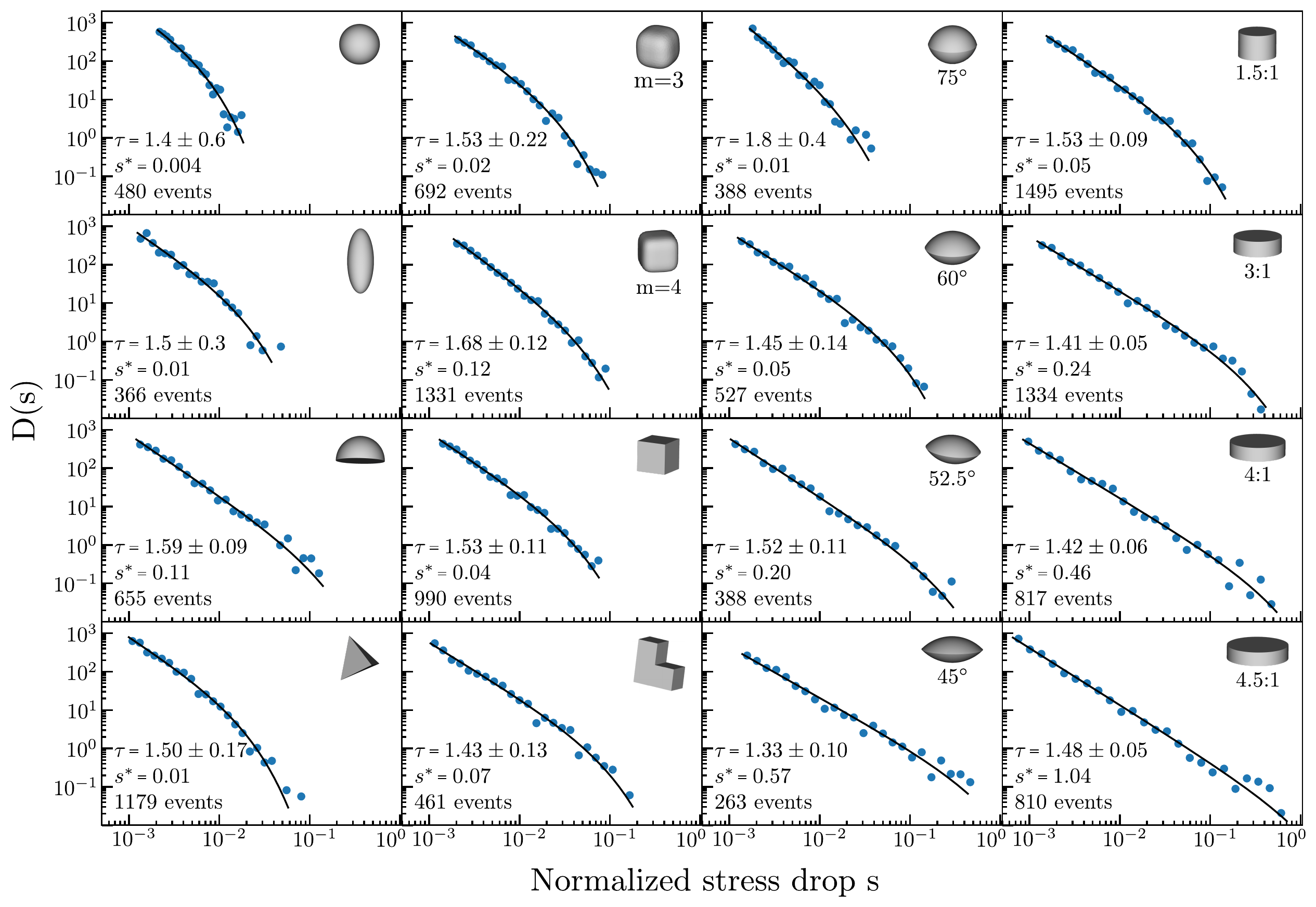}
	\caption{Stress drop magnitude distributions. 
    The drop distributions were fit by Maximum Likelihood Estimation to a truncated power law of the form $D(s)\sim s^{-\tau} exp(-s/s^*)$. 
    The values cited for the uncertainty on $\tau$ are $2\sigma$ for the marginal distribution, and only the best fit value of $s^*$ is listed. 
    All plots are over the same range. Data points are binned logarithmically (blue) after fitting (black). 
    The distributions are organized by (from left to right) the first column containing spheres, simple deviations from a sphere, and tetrahedra,
the second column cubes and simple deviations from a cube, the third column the lens family, and the fourth column the disk family.}
\end{figure*}

\subsection{Stress Fluctuation Statistics}
 
\begin{figure}
	\centering
	\includegraphics[width=1.0\linewidth]{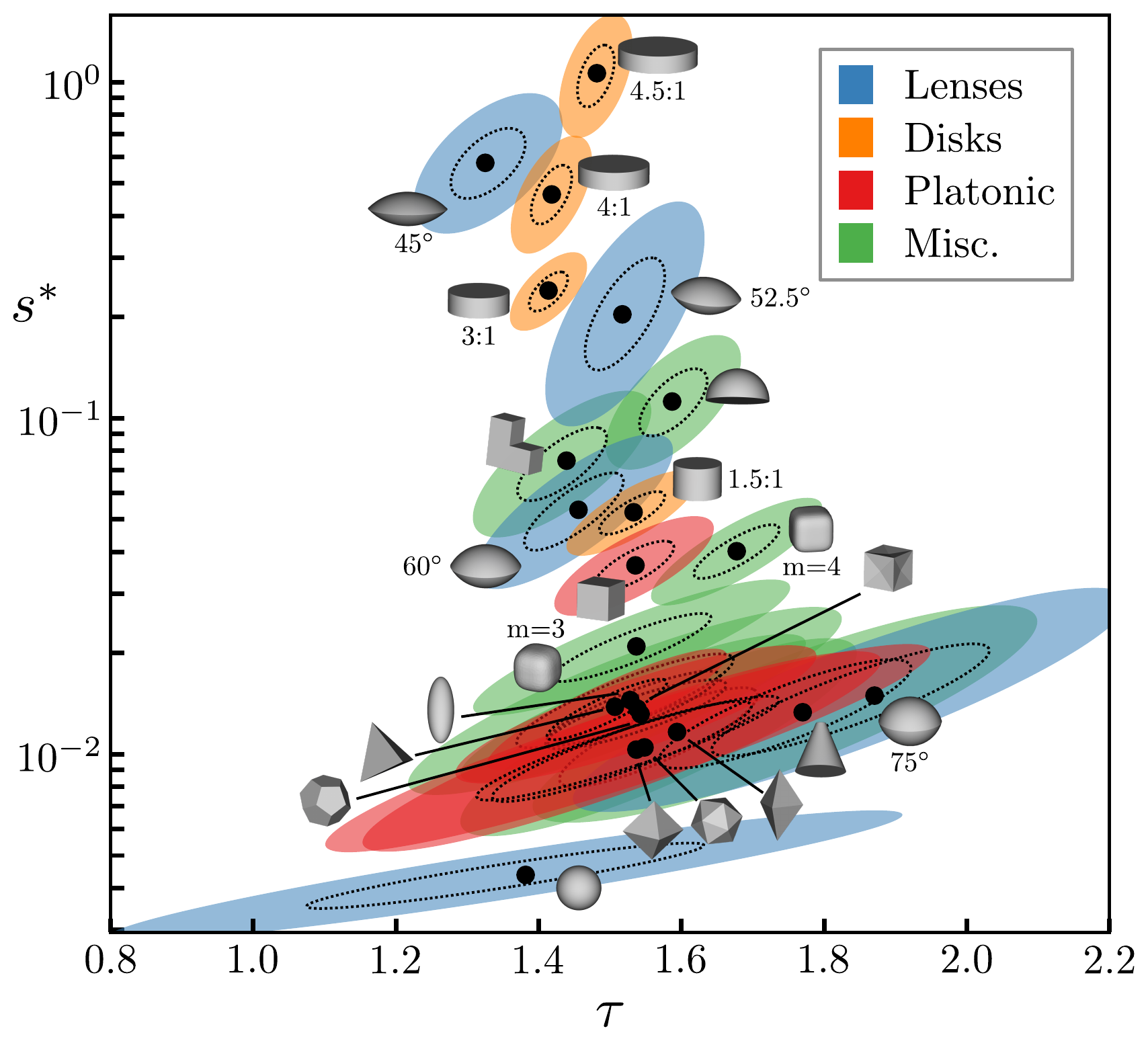}
	\caption{Parameter space for the truncated power law fits to $D(s)$. 
    The dotted (solid) ellipses around each best-fit $(\tau,s^*)$ are the $1\sigma$ ($2\sigma$) confidence regions as determined by the nonparametric bootstrap method.}
\end{figure}

We now turn to the statistics for individual stress drops $\Delta q$. 
In Fig. 5 we show the distribution $D(s)$ of normalized drop magnitudes $s=\Delta q/\bar{q}$ for a selection of the particle shapes tested. 
Here $\bar{q}$ is the average deviatoric stress in the critical state as discussed above within the context of Eq. 3. 
A first observation from Fig. 5 is that all distributions have roughly similar shape, with decreasing probability for increasing stress drop size and a cutoff beyond some maximum drop size. 
The lower end of the accessible range of $s$ in these distributions is given by the experimental noise floor, while the upper end corresponds to catastrophic events with drop magnitudes that are a significant fraction of the average stress. 

To extract more detailed, shape-dependent information, the distributions $D(s)$ were fitted numerically with Maximum Likelihood Estimation (MLE) \cite{ClausetPowerLaws2009} to a truncated power law $D(s)\sim s^{-\tau} exp(-s/s^*)$, as used in \cite{DenisovDahmenSchall2016}. Uncertainties on the two fit parameters, the power law exponent $\tau$ and the cutoff or characteristic stress scale $s^*$, were estimated using the nonparametric bootstrap method \cite{Bootstrap1993} with 1000 resamplings of the data. 
The drop distributions for some particle shapes, such as the spheres, have very small $s^*$ and thus no significant power law portion associated with $D(s)$, which means they could be well fit by a simple exponential. 
Others, such as for the flatter disks and lenses have significantly larger $s^*$ and offer nearly three decades in the scale-free part of their drop magnitude distribution, yielding a remarkably wide range from which one can extract the power law exponent. 

The best fit values for $\tau$ and $s^*$ are shown as black points in Fig. 6, along with the surrounding regions of uncertainty. 
These regions are the 1$\sigma$ and 2$\sigma$ ellipses for Gaussian fits to the bootstrap points in $(\tau, log \; s^*)$ space.

As with plotting $V$ against $\psi$ in Fig. 3, a plot of $s^*$ versus $\tau$ reveals trends for different particle shapes. 
More oblate shapes are seen to result in larger $s^*$, which means that compared to other shapes they can generate collapse-recovery events that involve a larger fraction of the average stress. 
However, $s^*$ does not correlate with a packing's shear strength, as seen from the similarly low $s^*$ values for spheres and bipyramids, shapes with very different $\psi$ (see Fig. 3). 
The cubes, supercubes, and divot cubes are all located near each other in Fig. 6, indicating that these shape variations do not have a major effect on the shape of $D(s)$, although there is a small shift in $s^*$ that is similar to the trend seen with volatility $V$. 

An important point that emerges from Fig. 6 is that $s^*$, the characteristic stress scale for the largest events, varies over two orders of magnitude while the ratio of system volume to particle volume remains fixed (except for minor variations in packing fraction). 
This clearly demonstrates that $s^*$ is not tied to system size. 
In fact, since in our experiments there are on the order of $10^2$ particles within a column cross section (fewer for the oblate shapes since they tend to settle and preferentially pack horizontally), a cutoff $s^*$ between $10^{-2}$ and $10^{-1}$, as seen in Figs. 5 and 6, is of the same order as if a single particle went from load-bearing member of the packing to unstressed rattler. 

From this alone we can infer that the power law regime in $D(s)$ is not due to large cascades of many-particle reconfigurations. 
Indeed, our x-ray imaging confirms that the vast majority of the measured stress drops corresponds to detectable shifts in the position of no more than a couple particles (Fig. 2). 
This suggests partial slips at the particle contacts, which relax only a portion of the supported force, as the elementary components of the scale-free cascades constituting stress drops, similar in nature to the partial stress drops thought to occur during earthquakes \cite{HeatonSlipPulse1990} and the partial stress drops incorporated in some mesoscale models\cite{DahmenSimpleAnalytic2011}.

\begin{figure}
	\centering
	\includegraphics[width=1.0\linewidth]{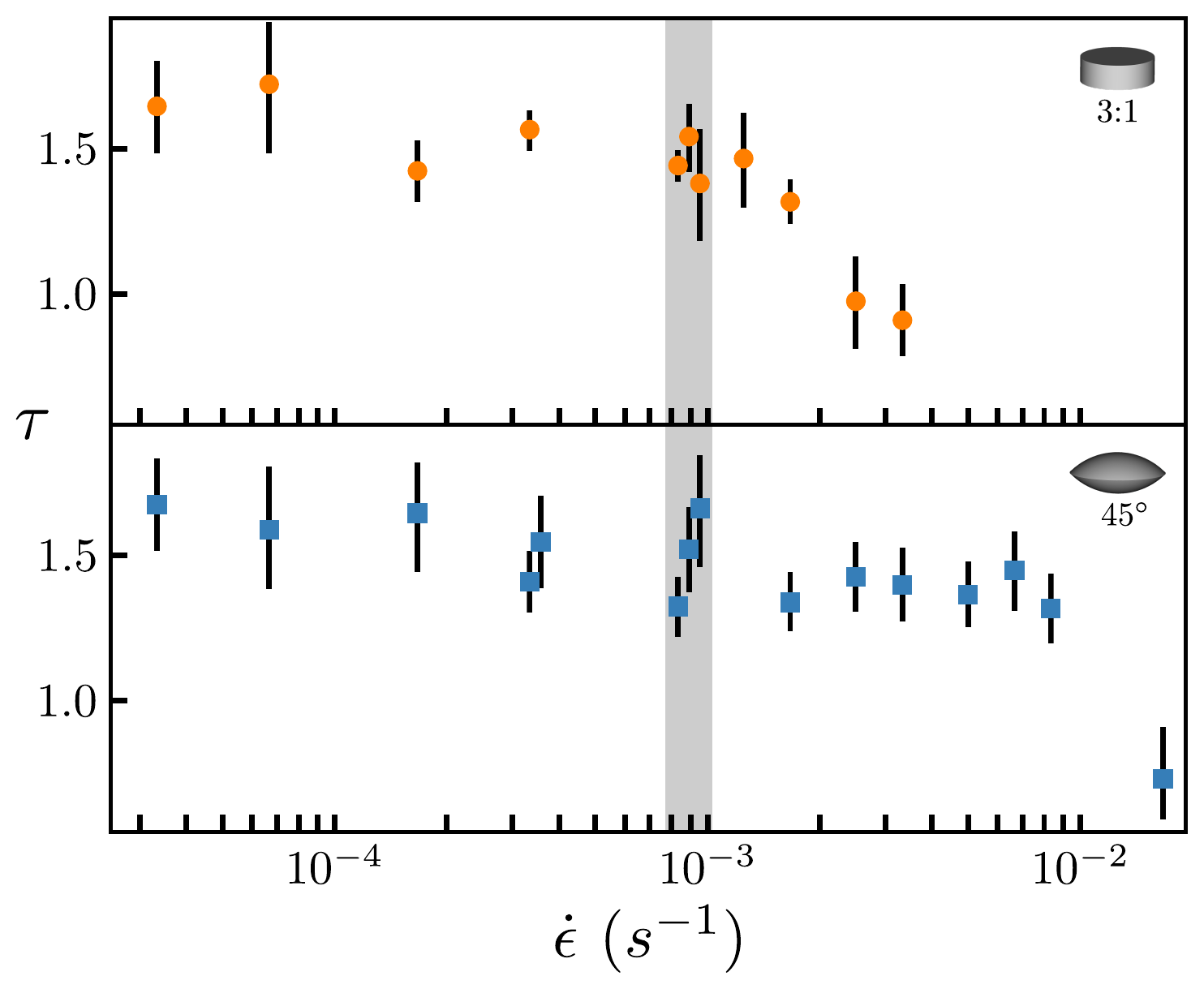}
	\caption{Strain rate dependence of $\tau$. Uncertainties are calculated as before with the bootstrap method. 
    The gray region contains the points which were taken at the same strain rate as the rest of the paper, $\dot{\epsilon}=8\times 10^{-4}$. 
    These points for both particle shapes and the $\dot{\epsilon}=3\times 10^{-4}$ data for the lenses are offset slightly to display multiple sets of runs taken at the same strain rate.}
\end{figure}

In light of the view that many plastically deforming systems, including granular materials, operate near a non-equilibrium critical phase transition\cite{WyartCriticality2015,DahmenSimpleAnalytic2011}, the value of $s^*$ increases with proximity to the critical point. 
The clear shape dependence shown in Fig. 6 indicates, then, that by varying particle shape we are able to create granular packings which plastically deform further from or closer to the point in system space where stress drop cascades are entirely scale free and correlations span the system.

Regarding the exponent $\tau$, clean trends are harder to isolate. Given the fit uncertainties, most shapes have $\tau$ in the range 1.3-1.7 and would be compatible with a value of 1.5. 
Recent mesoscale plasticity models have focused on the kernel used to describe stress redistribution after local yielding. 
The kernel used in \cite{DahmenSimpleAnalytic2011} resembles sandpile models where stress is offloaded isotropically to all neighbors, which lends itself to mean field treatment and a value of $\tau$ equal to 3/2.
Lin \textit{et al.} showed \cite{Wyart2014} that a modified kernel of quadrupolar form, found by Eshelby to be the response of an elastic continuum to localized relaxation\cite{Eshelby1957}, leads to $\tau \approx$ 1.3 in three dimensions. 
A tensorial approach to more accurately represent three dimensional stress and strain, also using a quadrupolar kernel, observed $\tau$=1.28 in simulations across a wide range of loading conditions \cite{TensorModelZapperi2017}. 
That the $\tau$ observed in the present experiments less closely matches the $\tau$ for models with a quadrupolar stress redistribution might indicate either that there is some effective degree of incipient shear banding in the experiments (though no fully formed shear bands were observed in any trials) or that the mesoscale granular characteristics of the aggregate material in our system do not reflect the continuum elastic response encapsulated by a quadrupolar kernel. 

The strain rate of compression has been shown to affect relaxation event magnitude distributions. 
Compression that is too rapid leads to event consolidation, which can be real \cite{DahmenRateDependence2003} or can be artificial due to poor data resolution\cite{DahmenTimeResolution2016}. 
In either case small events are missed and large events swell, leading to smaller power law exponents $\tau$. 
On the other extreme, when relaxation is present, compression of systems that is too slow can lead to dynamics that resemble self-organized criticality and have larger exponents $\tau$\cite{Papanikolaou2012}. 
Only the first, effects of rapid compression, are clearly observed in Fig. 7. 
Here we show the measured $\tau$ for the 3:1 aspect ratio disks and the 45$^\circ$ lenses across more than two orders of magnitude in strain rate $\dot{\epsilon}$. 
The exponent $\tau$ is found to be independent of strain rate up until a rate of compression faster than the $\dot{\epsilon}=8\times10^{-4}$ used in the data of Figures 3-6. 
The decrease in $\tau$ for both particle shapes happens when the compression is fast enough that the event duration timescale, independent of $\dot{\epsilon}$, is roughly half the timescale between events, a timescale which goes as $1/\dot{\epsilon}$. 
Indeed, the rate of detectable events for the 3:1 disks is about 5 times higher than for the 45$^\circ$ lenses, shown in Fig. 8 and qualitatively in the raw data of Fig. 1b. This factor of about 5 matches the difference in strain rates at which $\tau$ drops for each of these particle shapes in Fig. 7.

\section{Conclusions}

Our results demonstrate that deviations from spherical shape can have profound consequences on the character of stress fluctuations in plastically deforming granular systems at the mesoscale. 
For extracting shape-dependent trends in a model-free manner, we show that correlating a particle shape's propensity for generating large fluctuations (volatility) with its ability to resist shear (angle of internal friction) provides a powerful new tool (Fig. 3). 
This allows us to identify classes of shapes that differ in volatility by more than one order of magnitude despite providing similar resistance to shear. 
Specifically, large volatility $V$ is found in particle types that form amorphous packing configurations where nearest neighbor contacts tend to involve surfaces with large radius of curvature. 
Other features, such as sharp edges or corners, appear to be more important for enhancing the shear resistance. 
This is exemplified by the striking contrast between oblate lenses and flat disks (large $V$) on one hand and tetrahedra (small $V$) on the other. 

The cutoff event sizes $s^*$ that set the upper limit of the stress drop distributions for different particle types are also found to vary significantly (Fig. 6). 
As a comparison of Figs. 3 and 6 shows, trends in $s^*$ roughly mirror those in $V$, again highlighting the importance of face over edge contacts. 
Furthermore, with $s^*$ values for most particle shapes below $10^{-1}$ the fractional stress changes are small: given that there are on the order of 100 particles within a horizontal slice of the column, $s^*\approx$ 0.1 corresponds to complete loss of contact between no more than just a few particles. 
Supported by qualitative results from x-ray imaging, this suggests that the fluctuation statistics are dominated, in terms of frequency, by cascades of partial stress drop `microslips,' which only occasionally build into larger reorganization events involving groups of particles which shift significantly. 
Such microslips allow for small shifts in the relative particle positions across a contact, thereby changing the magnitude of the transmitted force without necessarily breaking the contact. 
In this context platy, oblate particles provide contact geometries well suited to accommodate many microslip events during a given relaxation event, which justifies their large $V$ and $s^*$ values. 

The values of the power law exponents $\tau$ we extracted from the stress drop distributions pose an intriguing problem for the current state of modeling for general amorphous plasticity. 
On the one hand, while the values are consistent with a mean field treatment of amorphous plasticity, which predicts $\tau=$ 3/2, we did not observe the accompanying shear localization (band formation) that would have been expected. 
On the other hand, while our results are less compatible with $\tau\approx$ 1.3  from models based on continuum elasticity with Eshelby's quadrupolar kernel, such a kernel arguably provides a more realistic description of the stress redistribution compared to isotropic mean field approaches. 

The fact that particle shape can change $s^*$ by over two orders of magnitude demonstrates the significant role of shape in setting the local interactions between constituent units and introduces a new means of controlling the distance a plastically deforming amorphous system operates from criticality.
Taken together with the results for $\tau$, these considerations imply a need to reexamine the mesoscale mechanisms driving amorphous plasticity. 
In this regard the map of $\tau$ and $s^*$ values in Fig. 6 can provide benchmark data for evaluating models.

Finally, our results also provide new insights for design applications of granular materials.  
Evolutionary algorithms have been employed to find optimal shapes for packing density \cite{MiskinMickey2014,RothTrimer2016} and strain-stiffening behavior \cite{MiskinWishbone2013}; the material parameters which can benefit most from shape optimization are clearly the ones for which particle shape plays a large role.
The sensitive dependence of plasticity behavior on particle shape shown in this work therefore reveals the shear strength and stress fluctuation magnitude to be especially amenable to optimization.  Figure 3 shows that with particle shape the two can be tailored independently of one another.
This opens the door for designed granular materials which enhance desirable properties such as self-healing capabilities, where failure events occur but the packing reforms and structural integrity remains, and energy dissipation.         

\textbf{Acknowledgements}
We thank Tom Witten, Sidney Nagel, Melody Lim, and Leah Roth for insightful discussions. This work was supported by the  National Science Foundation through grants CBET-1605075 (H.J.) and CBET-1336634 (K.D.). Additional support was provided by the Chicago MRSEC, funded by the NSF through grant DMR-1420709. K.M. acknowledges support from the Center for Hierarchical Materials Design (CHiMaD).
We thank the Kavli Institute of Theoretical Physics for hospitality at a workshop and for support through grant NSF PHY-1125915.

\section{Appendix A: Event count}
The rate of detectable events varies by up to a factor of six across shapes even when $s^*$ is approximately equal (Fig. 8).  As mentioned in the description of Fig. 7, when the timescale between events becomes comparable to the timescale of each event, the strain rate of compression is fast enough that $\tau$ of the drop magnitude distribution becomes distorted to smaller values.  

\begin{figure}
	\centering
	\includegraphics[width=1.0\linewidth]{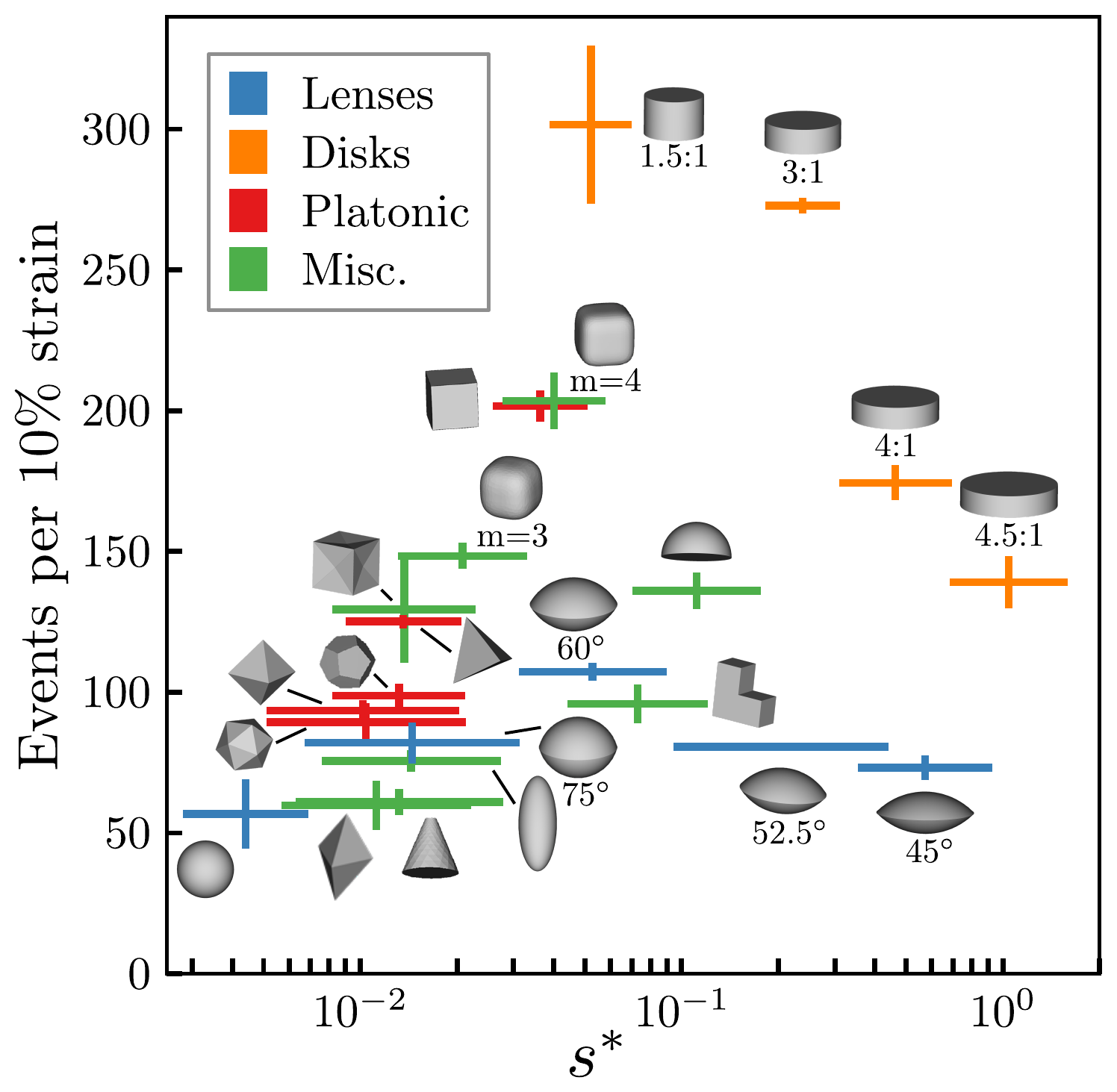}
	\caption{Detectable event rate versus slip event magnitude upper cutoff. Uncertainty for the event rate is calculated as the standard error of the mean count per run.}
\end{figure}

\end{document}